\documentclass[useAMS,usenatbib]{mn2e}
\usepackage[latin1]{inputenc}
\usepackage{fancyhdr}
\usepackage{multirow}
\usepackage{longtable}
\usepackage{color}
\usepackage{graphicx}
\usepackage{epsfig}
\usepackage{url}
\usepackage{xspace}

\bibpunct{(}{)}{;}{a}{}{,}
\usepackage{aas_macros} 

\title[Cluster mass bias from {\em Chandra}, {\em XMM}, \& {\em Planck}]{Reconciling \textit{Planck} cluster counts and cosmology? \textit{Chandra}/\textit{XMM} instrumental calibration and hydrostatic mass bias}

\author[H. Israel et al.]{Holger Israel$^{1,\ast}$,
   Gerrit Schellenberger$^{2}$,
   Jukka Nevalainen$^{3}$,
   \newauthor Richard Massey$^{1}$,
   and Thomas H. Reiprich$^{2}$\\
$^{1}$Institute for Computational Cosmology, Department of Physics, Durham University, South Road, Durham DH1 3LE, UK\\
$^{2}$Argelander-Institut f\"ur Astronomie, Auf dem H\"ugel 71, 53121 Bonn, Germany\\
$^{3}$Tartu Observatory, 61602 Toravere, Estonia\\
$^{\ast}$E-mail: {\tt holger.israel@durham.ac.uk}}

\begin{document}

\date{\today}

\pagerange{\pageref{firstpage}--\pageref{lastpage}} \pubyear{2014}

\maketitle

\label{firstpage}

\begin{abstract}
The mass of galaxy clusters can be inferred from the temperature of 
their X-ray emitting gas, $T_{\mathrm{X}}$. Their masses may be underestimated 
if it is assumed that the gas is in hydrostatic equilibrium, by an amount 
$b^{\mathrm{hyd}}\!\sim\!(20\pm10)$\% suggested by simulations.
We have previously found consistency between a sample of observed 
\textit{Chandra} X-ray masses and independent weak lensing measurements.
Unfortunately, uncertainties in the instrumental calibration of {\em Chandra} 
and {\em XMM-Newton} observatories mean that they measure different 
temperatures for the same gas.
In this paper, we translate that relative instrumental bias into mass bias, 
and infer that \textit{XMM-Newton} masses of 
$\sim\!10^{14}\,\mbox{M}_{\odot}$ ($\ga\!5\cdot10^{14}\,\mbox{M}_{\odot}$) 
clusters are unbiased ($\sim\!35$\% lower) compared to WL masses.
For massive clusters, \textit{Chandra}'s calibration may thus be more accurate.
The opposite appears to be true at the low mass end. 
We observe the mass bias to increase with cluster mass, but presence of 
Eddington bias precludes firm conclusions at this stage.
Nevertheless, the systematic \textit{Chandra} -- \textit{XMM-Newton} difference   
is important because {\em Planck}'s detections of massive clusters via the 
Sunyaev-Zeldovich (SZ) effect are calibrated via {\em XMM-Newton} observations.
The number of detected SZ clusters are inconsistent with {\em Planck}'s 
cosmological measurements of the primary Cosmic Microwave Background (CMB).
Given the \textit{Planck} cluster masses, if an (unlikely) uncorrected 
$\sim\!20$\% calibration bias existed, this tension would be eased, but not resolved.
\end{abstract}

\begin{keywords}
Galaxies: clusters: general  -- Cosmology: observations -- Gravitational lensing -- X-rays: galaxies: clusters
\end{keywords}

\section{Introduction}
The number of Sunyaev-Zeldovich (SZ) clusters detected with \textit{Planck} 
above a certain mass threshold \citep[P13XX]{2013arXiv1303.5080P} falls short
of the tally expected from the \textit{Planck} primary cosmic microwave 
background (CMB) constraints on cosmology \citep[P13XVI]{2013arXiv1303.5076P}.
Several possible explanations have been brought forward, such as incorrect 
assumptions about the cluster mass function (P13XX) or modified cosmologies
including massive neutrinos and a shift in the Hubble parameter
\citep[e.g., P13XX,][]{2013JCAP...10..044H,2014PhRvL.112e1303B,2014arXiv1407.4516M,2014arXiv1407.8338C}.
Another hypothesis is that hydrostatic cluster masses, inferred from X-ray 
observations of the intra-cluster medium (ICM), yielded 
only $\sim\!60$~\% of the true cluster mass.
Hydrodynamic cluster simulations commonly find the hydrostatic assumption to 
retrieve only $\sim\!70$--$90$~\% of the true cluster mass, i.e.
$M^{\mathrm{HE}}\!=\!(1-b_{\mathrm{lin}}^{\mathrm{hyd}}) M^{\mathrm{true}}$
with a hydrostatic mass bias $b_{\mathrm{lin}}^{\mathrm{hyd}}\!=\!0.1$--$0.3$ 
\citep[e.g.,][]{2007ApJ...668....1N,2010A&A...510A..76L,2012MNRAS.422.1999K,2012NJPh...14e5018R,2014MNRAS.441.1270L,2014arXiv1407.7040S}.

The validity of the assumption of hydrostatic equilibrium can potentially be 
addressed by comparing to weak gravitational lensing (WL) mass measurements, 
which are independent and free from assumptions of the state of the gas.
Noticing a considerable overlap between the \textit{XMM-Newton} sample of 
P13XX and the \textit{Weighing the Giants} WL survey 
\citep{2014MNRAS.439....2V,2014MNRAS.439...28K,2014MNRAS.439...48A},
\citet[vdL14]{2014MNRAS.443.1973V} measured 
$\langle M^{\mathrm{Pl}}\!/M^{\mathrm{wl}}\rangle\!=\!0.688\pm0.072$ 
for the most massive ($>\!6\cdot10^{14}\,\mbox{M}_{\odot}$) clusters.  
If interpreted as a hydrostatic mass bias, this value
$b_{\mathrm{lin}}\!\approx\!0.3$ falls short of the 
$b_{\mathrm{lin}}\!\approx\!0.4$ necessary to reconcile P13XX with P13XVI,
confirming the \textit{Planck} cluster mass discrepancy.

Conversely, \citet[I14]{2014A&A...564A.129I} found no significant mass bias
when comparing WL estimates to \textit{Chandra}-based hydrostatic masses. 
For high-mass clusters
($10^{14.5}\,\mbox{M}_{\odot}\!\!<\!\!M^{\mathrm{wl}}_{500}\!\!<\!\!10^{15}\,\mbox{M}_{\odot}$),
the bias $b_{\mathrm{log}}\!\!=\!\!-0.10_{-0.15}^{+0.17}$, 
is consistent with the expectation based on simulations, although with large 
uncertainties due to small number statistics. 

An alternative hypothesis is that at least one of the two X-ray observatories
is imperfectly calibrated. Indeed, difficulties modelling their 
(energy-dependent) effective collecting area \citep{2013arXiv1305.4480G} lead 
to uncertainty in measurements of the ICM temperature, $T_{\mathrm{X}}$.
Direct comparisons have shown that \textit{Chandra} measures significantly 
higher $T_{\mathrm{X}}$ than \textit{XMM-Newton} for the same clusters 
\citep[e.g.][]{2010A&A...523A..22N}, and that significant differences even 
exist between the \textit{XMM-Newton} instruments 
\citep[][S14]{2014arXiv1404.7130S}.
S14 propagated this difference to a change in the inferred cosmological matter
density $\Omega_{\mathrm{m}}$ and power spectrum normalisation $\sigma_{8}$.
They concluded that the temperature calibration alone is insufficient to 
explain the discrepancy between P13XVI and P13XX.

In this paper, we simultaneously examine the hydrostatic bias and 
\textit{XMM-Newton}/\textit{Chandra} instrument calibration, aiming to find a
solution for the cosmological discrepancy. We extend S14 by comparing 
measurements in an X-ray--selected cluster sample with independent WL masses.
In Section~\ref{sec:res}, we re-evaluate I14's measurements of the mass bias 
between \textit{Chandra} hydrostatic and WL masses, and emulate 
\textit{XMM-Newton} results based on S14's cross-calibration. Noting that the 
P13XX calibration relies on \textit{XMM-Newton}, in Section~\ref{sec:conclu}, 
we assess the degree to which X-ray temperature calibration could be 
responsible for the P13XVI--P13XX discrepancy.
We conclude in Section~\ref{sec:conclusion}.

\section{Recalibrating the \emph{400d} survey to \textbfit{XMM-Newton} temperatures} \label{sec:res}

\subsection{Hydrostatic mass bias from the \emph{400d} cluster cosmology survey}

I14 recently compared WL masses to \textit{Chandra}-based X-ray mass estimates
for eight clusters drawn from the \emph{400d} cosmology cluster sample.
The \emph{400d} cosmology sample selects X-ray luminous clusters at 
$0.35\!<\!z\!<\!0.90$ from the serendipitous \emph{400d} \textit{Rosat} 
cluster catalogue \citep{2007ApJS..172..561B}. \textit{Chandra} data for these 
clusters were subsequently employed to constrain cosmological parameters
via the cluster mass function \citep{2009ApJ...692.1033V,2009ApJ...692.1060V}.
The \emph{400d} WL survey follows up the cosmology cluster sample, in order to 
test the mass calibration of V09a,b with independent mass estimates.
The methodology and first results of the ongoing \emph{400d} WL survey were 
reported in \citet{2010A&A...520A..58I,2012A&A...546A..79I}.
We refer the interested reader to these papers for details.
Weak lensing masses used in this paper make use of the 
\citet{2013ApJ...766...32B} mass--concentration relation.

Hydrostatic masses in I14 were derived from the V09a \textit{Chandra} ICM
density profiles $\rho_{\mathrm{g}}$ using the \citet{2006ApJ...640..691V} 
parametrisation, and temperatures $T_{\mathrm{X}}(r)\!=\!T_{\mathrm{CXO}}(r)$.
The empirical \citet{2013SSRv..177..195R} relation was used
to derive a temperature profile
\begin{equation} \label{eq:tprof}
T_{\mathrm{X}}(r)\!=\!T_{\mathrm{X}}\left(1.19-0.84r/r_{200}\right) 
\end{equation}
from a cluster-averaged value $T_{\mathrm{X}}$ and I14 WL radius $r_{200}$.
This relation was determined and can be used in the range 
$0.3\,r_{200}\!<\!r\!<\!1.15\,r_{200}$. We then compute
\begin{equation} \label{eq:hse}
M^{\mathrm{HE}}(r)\!=\!\frac{-k_{\mathrm{B}}T_{\mathrm{X}}(r)\,r}{\mu m_{\mathrm{p}}G}
\left(\frac{\mathrm{d}\ln \rho_{\mathrm{g}}(r)}{\mathrm{d}\ln r} + 
\frac{\mathrm{d}\ln T_{\mathrm{X}}(r)}{\mathrm{d}\ln r}\right),
\end{equation}
with $k_{\mathrm{B}}$ the Boltzmann constant, $\mu\!=\!0.5954$ the 
mean molecular mass of the ICM, $m_{\mathrm{p}}$ the proton mass, and
$G$ the gravitational constant. The resulting cumulative mass profile was 
evaluated at $r_{500}$ taken from WL. Uncertainties on $T_{\mathrm{CXO}}$ and 
$r_{500}^{\mathrm{wl}}$ were propagated into an uncertainty on 
$M^{\mathrm{hyd}}_{500}(r_{500}^{\mathrm{wl}})$.

\subsection{Monte Carlo analysis} \label{sec:mc}

We adopt a Monte Carlo approach to derive hydrostatic masses.
In our scheme, the V09a cluster-averaged temperature $T_{\mathrm{CXO}}$, the
square of the \citet[I12]{2012A&A...546A..79I} WL cluster radius 
$(r_{500}^{\mathrm{WL}})^{2}$, and the slope, normalisation, and intrinsic 
scatter of the S14 \textit{Chandra} $\leftrightarrow$ \textit{XMM-Newton} 
calibration relation are sampled from their Gaussian distributed probability 
densities. We point out that we model each of our clusters independently.
By choosing $(r_{500}^{\mathrm{WL}})^{2}$, whose I12 measurements we 
empirically find to follow a normal distribution, we are able to easily 
reproduce the asymmetric uncertainties in $r_{500}$, improving our treatment 
from I14. Using $10^{6}$ Monte Carlo realisations, we excellently recover the 
I12 WL masses. Through the use of $r_{500}^{\mathrm{WL}}$ in 
Eq.~(\ref{eq:tprof}), the updated account of its asymmetric uncertainties
results in slightly lower \textit{Chandra} hydrostatic masses.
Compared to I14, \textit{Chandra} hydrostatic masses are lower by an average
$(1.2\pm0.3)$~\% (compare Table~\ref{tab:res} to Table~2 of I14).
Our new Monte Carlo technique leads to smaller uncertainties in the hydrostatic
masses compared to the conservative combination of uncertainties in 
$T_{\mathrm{CXO}}$ and $r_{500}^{\mathrm{WL}}$ that was employed by I14.

\subsection{Pseudo-\textit{XMM-Newton} temperatures for the \emph{400d} clusters} \label{sec:txmm}

\begin{figure}
 \includegraphics[width=8cm]{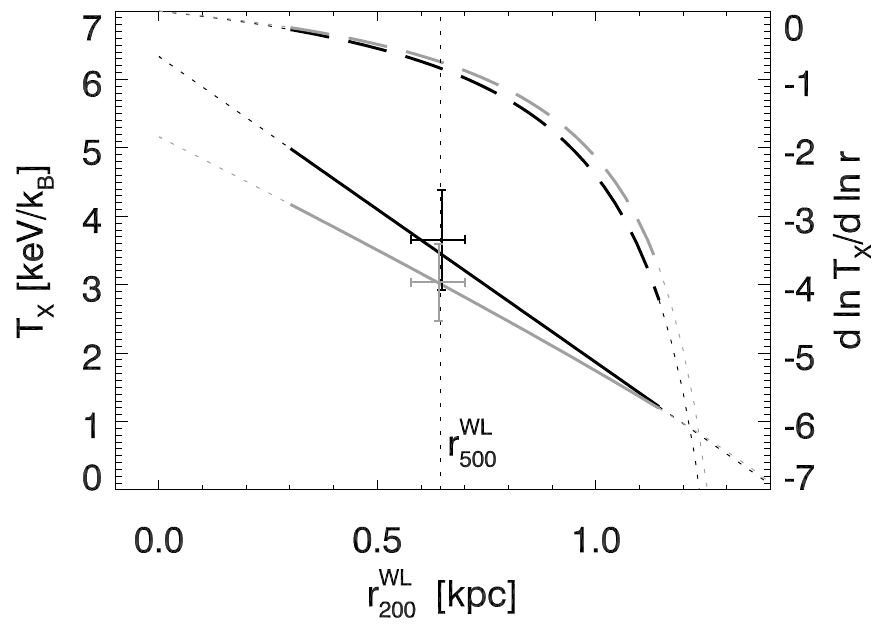}
 \caption{The effect of re-calibration on the temperature profile.
The black solid line shows the \citet{2013SSRv..177..195R} \textit{Chandra} 
temperature profile we assume for CL\,0030+2618. By applying 
Eq.~(\ref{eq:s14}), which is linear in $\log{T_{\mathrm{X}}}$, to each datum
of the profile, we derive the grey solid pseudo-\textit{XMM-Newton} profile, 
which is slightly curved, but still close to the \citet{2013SSRv..177..195R} 
form. As indicated by the vertical line, $r_{500}$ lies safely within the range
(bold lines) in which the \citet{2013SSRv..177..195R} profile can be used.
Long-dashed curves denote the logarithmic derivatives. For the sake of clarity,
uncertainties are only shown at $r_{500}$.}
\label{fig:temp}
\end{figure}
The International Astronomical Consortium for High Energy Calibration (IACHEC)
has tasked itself with improving (cross-)calibrations of X-ray satellite 
observatories \citep{2013arXiv1305.4480G}.
In this context, \citet[S14]{2014arXiv1404.7130S}
published a detailed comparison of \textit{Chandra} and \textit{XMM-Newton}
temperatures for the HIFLUGCS sample of $64$ high-flux
local clusters, fitting spectra in the same radial and energy ranges. 
They not only confirmed earlier studies \citep[e.g.][]{2010A&A...523A..22N}
that \textit{Chandra} yields significantly higher
$T_{\mathrm{X}}$ than \textit{XMM-Newton}, but also find significant 
differences between the \textit{XMM-Newton} instruments. These temperature 
differences are most pronounced at the highest plasma temperatures and can 
best be explained as calibration uncertainties on effective area.

For the \emph{400d} cluster sample, we translate ICM temperatures measured with
\textit{Chandra}, $T_{\mathrm{CXO}}$, to pseudo-\textit{XMM-Newton} 
temperatures by applying the S14 conversion formula between ACIS and the 
combined \textit{XMM-Newton} instruments for $0.7$--$7$ keV energy range:
\begin{equation} \label{eq:s14}
\log{\left(\frac{k_{\mathrm{B}}T_{\mathrm{XMM}}}{1\,\mbox{keV}}\right)} = 
A\cdot\log{\left(\frac{k_{\mathrm{B}}T_{\mathrm{CXO}}}{1\,\mbox{keV}}\right)} + B\quad.
\end{equation}
Both the calibration of an X-ray instrument and our knowledge about it evolve
with time. S14 assume calibrations as of December 2012 (\textit{Chandra} 
Calibration Database v4.2), while V09a used the unchanged 
\citet{2005ApJ...628..655V} calibration procedure. This is no Calibration 
Database calibration, but at the time of observation v3.1 was in place.
Therefore, we apply the following steps to derive 
pseudo-\textit{XMM-Newton} temperature profiles: 

\begin{description}
 \item[1.] We transform the V09a $T_{\mathrm{CXO}}$ from the energy range of
$0.6$--$10$ keV to $0.7$--$7$ keV, by applying a correction
\begin{equation}
\log{\left(\frac{k_{\mathrm{B}}T_{\mathrm{CXO}}^{(0.7-7)}}{1\,\mbox{keV}}\right)} = 
A_{0}\cdot\log{\left(\frac{k_{\mathrm{B}}T_{\mathrm{CXO}}^{(0.6-10)}}
{1\,\mbox{keV}}\right)} + B_{0}
\end{equation}
with $A_{0}\!=\!1.0027\pm0.0018$ and $B_{0}\!=\!-0.0008\pm0.0013$
derived from fitting the \textit{Chandra} temperatures of the HIFLUGCS sample
in the two spectral ranges, in analogy to S14. 
This raises the $T_{\mathrm{CXO}}$ values by $0.1$~\% to $0.3$~\%.

\item[2.]
Using the time\-stamp correction for $T_{\mathrm{X}}$ between different 
Calibration Databases \citep{2010ApJ...721..653R}, derived for the 
$0.7$--$7$ keV band, we convert the V09a temperatures to the one used by S14 
(version 4.2). From Eq.~(23) of \citet{2010ApJ...721..653R}, we take a factor
of $T_{\mathrm{CXO,3.1}}/T_{\mathrm{CXO,4.2}}\!=\!1.06\pm0.05$.

\item[3.]
For each of the $10^{6}$ Monte Carlo realisations, we compute the 
\textit{Chandra} temperature profile following Eq.~(\ref{eq:tprof}). 
The black solid line in Fig.~\ref{fig:temp} shows an example (CL~0030+2618).

\item[4.] 
Finally, we perform the transformation (Eq.~\ref{eq:s14}) between 
\textit{Chandra} and the combined \textit{XMM-Newton} instruments, in the 
$0.7$--$7$ keV energy range. The best-fit parameters taken from S14 are 
$A\!=\!0.889_{-0.003}^{+0.005}$ and $B\!=\!0.000\pm0.004$.
This transformation is applied to every datum of the temperature profile.
As the grey solid line in Fig.~\ref{fig:temp} shows, the re-calibration 
introduces a slight curvature, because Eq.~(\ref{eq:s14}) is linear in 
$\log{T_{\mathrm{X}}}$ rather than in $T_{\mathrm{X}}$. Given the measurement 
uncertainties, the resulting departure from the form of Eq.~(\ref{eq:tprof})
is insignificant.
\end{description} 

By applying this conversion, we emulate what ICM temperatures 
would have been obtained for the \emph{400d} clusters, had they been 
inferred from both the Metal Oxide Semi-conductor (MOS) and  the pn-CCD (PN) 
instruments (collectively, the \textit{XMM-Newton European Photon Imaging 
Camera}, EPIC) instead of \textit{Chandra}'s Advanced CCD Imaging Spectrometer
(ACIS). We denote the resulting temperatures $T_{\mathrm{xmm}}$, with the 
lowercase indicating that they are converted quantities, not actual 
\textit{XMM-Newton} measurements.

For the eight I14 clusters, whose 
$\langle T_{\mathrm{CXO}}\rangle\!=\!4.4\,\mbox{keV}\!/k_{\mathrm{B}}$ 
is representative of the full \emph{400d} cosmology sample, we measure 
$\langle T_{\mathrm{xmm}}/T_{\mathrm{CXO}}\rangle\!=\!0.81\pm0.01$,
using the V09a cluster-averaged temperatures.
At $r_{500}$, measured from weak lensing, the ratio is 
$\langle T_{\mathrm{xmm}}/T_{\mathrm{CXO}}\rangle\!=\!0.85\pm0.01$. 
This ratio is closer to $1$ because 
$T_{\mathrm{X}}(r_{500})\!<\!\langle T_{\mathrm{X}}\rangle$ and the 
cross-calibration differences are smaller for lower $T_{\mathrm{X}}$ 
according to S14.

\subsection{Pseudo-\textit{XMM-Newton} hydrostatic masses} \label{sec:mxmm}
Within our Monte Carlo scheme, we re-derive hydrostatic masses by 
inserting the pseudo-\textit{XMM-Newton} profiles $T_{\mathrm{xmm}}(r)$ and 
their values at $r_{500}$ \color{black} into Eq.~(\ref{eq:hse}),
thus accounting for the nonlinear nature of Eq.~(\ref{eq:s14}).

Differences in the effective area normalisation between \textit{Chandra} and 
\textit{XMM-Newton} also affect the measured gas mass $M_{\mathrm{gas}}$ and
hydrostatic mass via the calibration of the flux $S$. 
As $M_{\mathrm{gas}}\!\propto\!\sqrt{S}$, the $5$~\% flux difference 
for the full energy range in \citet{2010A&A...523A..22N} correspond to $2$~\% 
uncertainty in $M_{\mathrm{gas}}$. We account for this effect by rescaling
the pseudo-\textit{XMM-Newton} masses by $0.98$.

\begin{figure}
 \includegraphics[width=8cm]{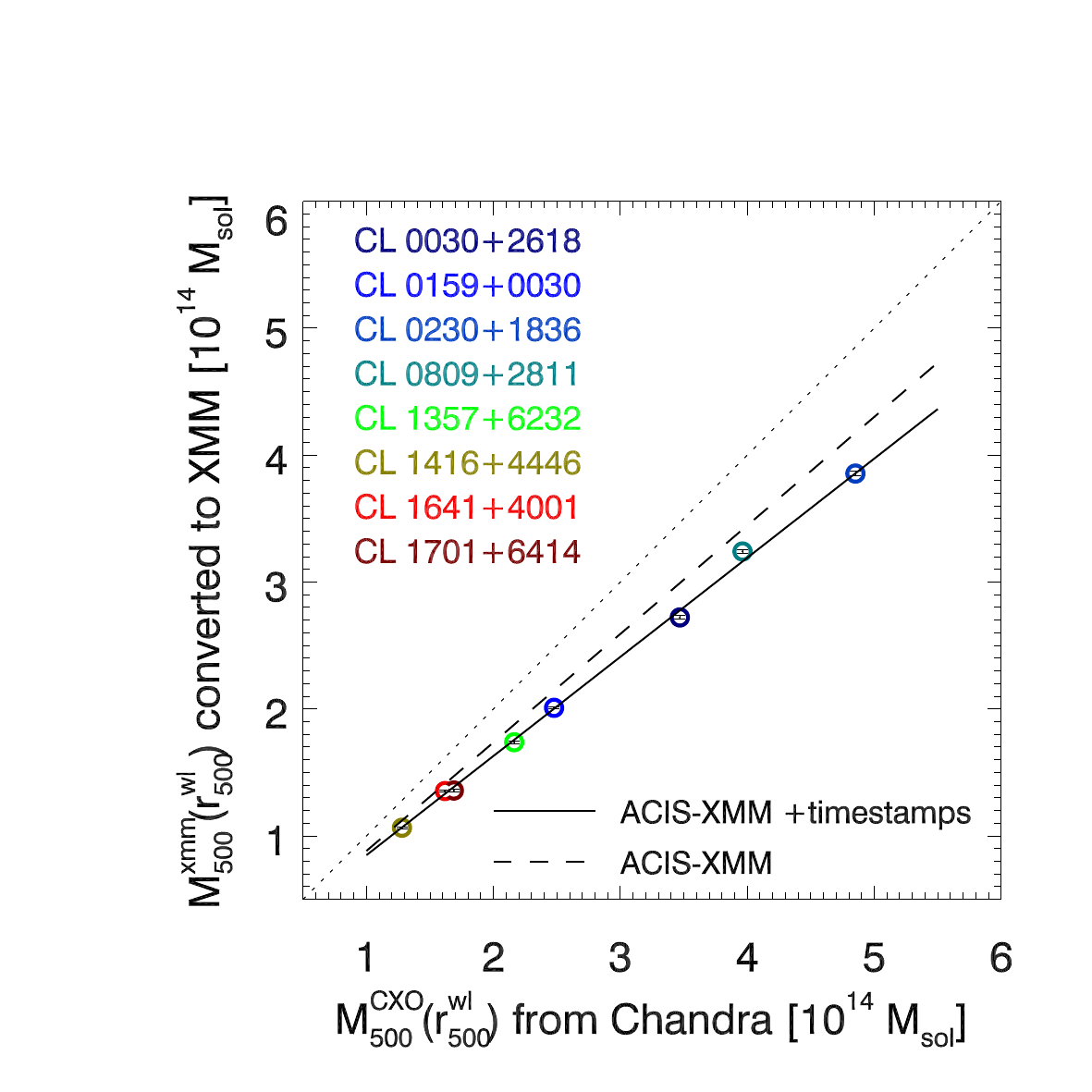}
 \caption{Mass estimates $M^{\mathrm{xmm}}_{500}$ derived from 
pseudo-\textit{XMM-Newton} temperatures and assuming hydrostatic equilibrium 
as a function of masses $M^{\mathrm{CXO}}_{500}$ derived from ICM temperatures 
observed by \textit{Chandra}. Error bars inscribed in the symbols denote the
uncertainty in $M^{\mathrm{xmm}}_{500}$ due to the uncertainties in the 
ACIS--combined XMM and time\-stamp conversions. For 
illustrative purposes, the time\-stamp correction is not applied to the 
$M^{\mathrm{CXO}}_{500}$, but its inverse to the $M^{\mathrm{xmm}}_{500}$.
The solid line marks the linear best fit. A dashed line marks the best-fit 
relation when the different \textit{Chandra} calibration time\-stamps are not
taken into account.
For the latter case, data points are not shown for the sake of clarity.}
  \label{fig:xmmchandra}
\end{figure}
As expected for lower input temperatures and flatter $T_{\mathrm{X}}$ 
gradients, we find the resulting pseudo-\textit{XMM-Newton} hydrostatic masses
for all clusters to be lower than the \textit{Chandra}-measured values
(Fig.~\ref{fig:xmmchandra}). 
We point out that in Fig.~\ref{fig:xmmchandra}, we do not apply the time\-stamp 
correction to the $T_{\mathrm{CXO}}$, to highlight the combined effect of both 
corrections. The relative difference in masses is strongest for the hottest 
clusters, for which the S14 conversion results in the largest change. 
Because the I14 sample exhibits a limited $T_{\mathrm{X}}$ range of 
$3$--$6\,\mbox{keV}$, the relative change of the temperatures varies less than
$5$~\%. Consequently, the two sets of hydrostatic masses are well fit by a 
linear relation (solid line in Fig.~\ref{fig:xmmchandra}):
\begin{equation} 
\frac{M^{\mathrm{xmm}}_{500}}{10^{14}\mbox{M}_{\odot}} = P\cdot
\frac{M^{\mathrm{CXO}}_{500}}{10^{14}\mbox{M}_{\odot}} + Q 
\label{eq:xmmcxo}
\end{equation} 
with $P\!=\!0.783\!\pm\!0.007$ and $Q\!=\!0.062\!\pm\!0.015$ that captures the
dependence of the \textit{Chandra}--\textit{XMM-Newton} disagreement on the 
measured mass itself. As a sample average and standard error, we find
$1\!-\!b_{\mathrm{lin}}^{\mathrm{xcal}}\!=\!1\!-\!\langle M^{\mathrm{xmm}}_{500}\!/
M^{\mathrm{CXO}}_{500}\rangle\!=\!0.81\pm0.01$.
The difference between this number and
$1\!-\!\langle T_{\mathrm{xmm}}(r_{500}^{\mathrm{wl}})/T_{\mathrm{CXO}}(r_{500}^{\mathrm{wl}})\rangle\!=\!0.15\pm0.01$ 
can be traced back to the additional factor of
$T_{\mathrm{X}}\left(\frac{\mathrm{d}\ln T_{\mathrm{X}}(r)}{\mathrm{d}\ln r}\right)$
in Eq.~(\ref{eq:hse}).

\subsection{Stronger WL mass bias for pseudo-\textit{XMM-Newton} masses} \label{sec:mbias}
\begin{figure}
 \includegraphics[width=8cm]{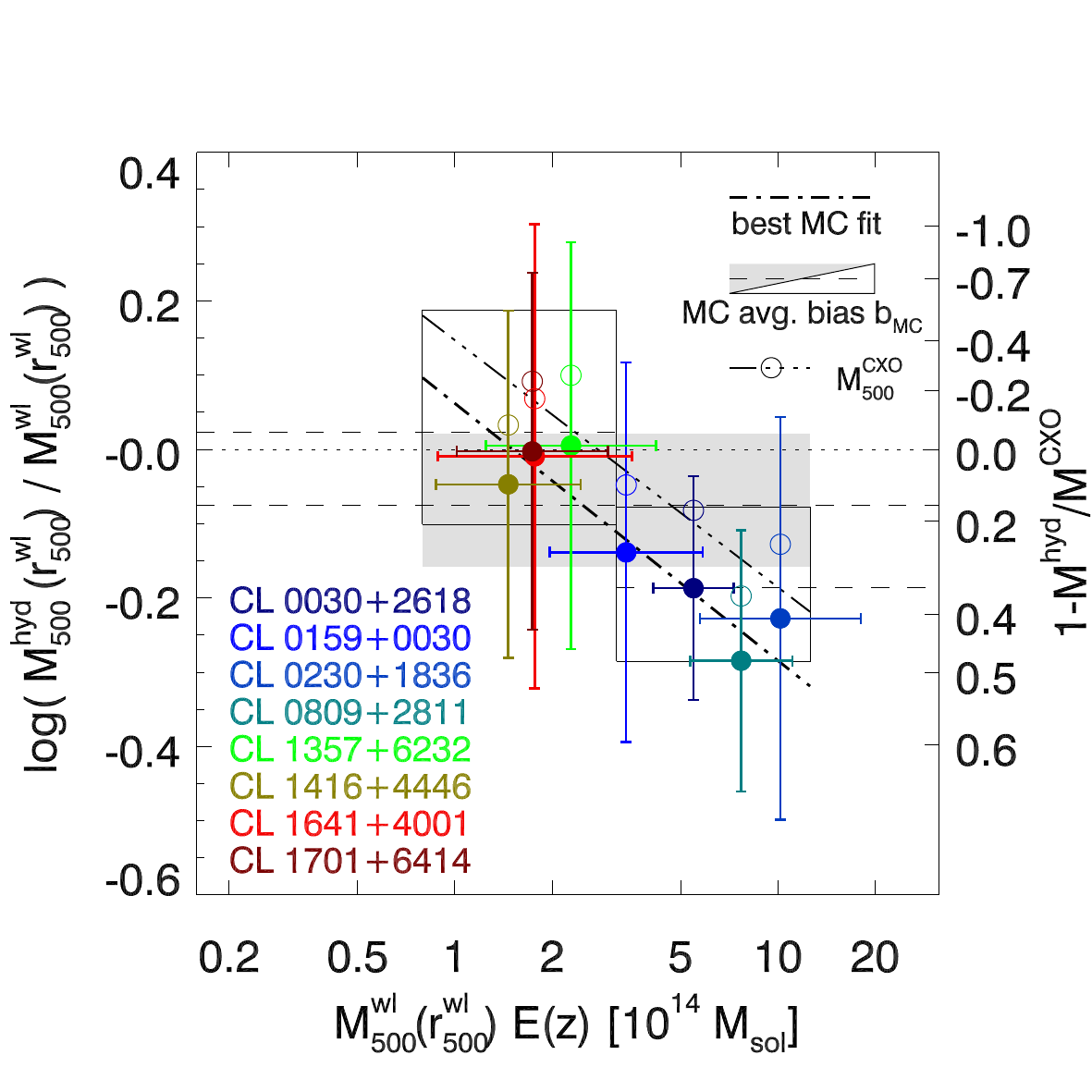}
 \caption{Ratio between the pseudo-\textit{XMM-Newton} hydrostatic mass 
$M_{500}^{\mathrm{xmm}}$, with time\-stamp correction, and the I14 WL mass 
$M_{500}^{\mathrm{wl}}$ as a function of $M_{500}^{\mathrm{wl}}$.
Short-dashed lines and light grey shading denote the logarithmic bias
$b_{\mathrm{log}}\!=\!\langle\log M^{\mathrm{xmm}}\!-\!\log M^{\mathrm{wl}}\rangle$ 
obtained from averaging over Monte Carlo realisations. We also show 
$b_{\mathrm{log}}$ for the low-$M^{\mathrm{wl}}$ and high-$M^{\mathrm{wl}}$ 
clusters separately, with the $1\sigma$ uncertainties presented as boxes, 
for sake of clarity. As a visual aid, a dot-dashed line depicts the 
Monte Carlo best-fit of $\log{(M^{\mathrm{xmm}}/M^{\mathrm{wl}})}$ as a 
function of $M^{\mathrm{wl}}$. Empty symbols and the triple-dot-dashed line
denote the  $M_{500}^{\mathrm{CXO}}$ case. Compare to Fig.~2A in I14.}
  \label{fig:bias}
\end{figure}
\begin{table*}
 \caption{Observed mass bias in the I14 sample, for several choices of X-ray 
masses. Columns 2 and 3 give the slope $P$ and intercept $Q$ of the general
best-fit relation (Eq.~\ref{eq:xmmcxo}) between \textit{Chandra} and 
\textit{XMM-Newton} masses. Column 4 shows the X-ray calibration bias, 
i.e.\ the mean and standard error of 
$\langle M^{\mathrm{xmm}}_{500}/M^{\mathrm{CXO,I14}}_{500}\rangle$. 
Columns 5 and 6 show the apparent bias with respect to the \textit{Chandra} 
masses, averaged over Monte Carlo simulations for all clusters
($b_{\mathrm{log}}\!=\!\langle\log{M^{\mathrm{xmm}}_{500}}\!-\!\log{M^{\mathrm{CXO,I14}}_{500}}\rangle$)
 and for the $M_{500}^{\mathrm{wl}}\!\geq\!10^{14.5}\,\mbox{M}_{\odot}$ bin 
($b_{\mathrm{log,H}}$). The final column measures the mass-dependent mass bias
as the difference $\Delta b_{\mathrm{log}}^{\mathrm{H-L}}$ between 
$b_{\mathrm{log}}$ for the high- and low-mass clusters.}
 \renewcommand\tabcolsep{3pt}
  \begin{tabular}{ccccccc}\hline\hline
Hydrostatic mass & $P$ & $Q$ & $b^{\mathrm{xcal}}_{\mathrm{lin}}$ & $b_{\mathrm{log}}$ &
   $b_{\mathrm{log,H}}$ & $\Delta b_{\mathrm{log}}^{\mathrm{H-L}}$ \\ \hline
$M_{500}^{\mathrm{CXO}}$, new Monte Carlo & $1$ & $0$ & $0$ & $0.02_{-0.08}^{+0.10}$ &
   $-0.09_{-0.10}^{+0.11}$ & $-0.20_{-0.16}^{+0.20}$ \\
$M_{500}^{\mathrm{CXO}}$, incl.\ time\-stamp correction & $0.946\pm0.009$ & $-0.002\pm0.020$ & $0.06\pm0.00$ & 
   $-0.01_{-0.09}^{+0.10}$ & $-0.11\pm0.11$ & $-0.20_{-0.16}^{+0.20}$\\
$M_{500}^{\mathrm{xmm}}$, full conversion & $0.783\pm0.007$ & $0.062\pm0.015$ & $0.19\pm0.01$ &
   $-0.08_{-0.08}^{+0.10}$ & $-0.19_{-0.10}^{+0.11}$ & $-0.21_{-0.16}^{+0.20}$\\
$M_{500}^{\mathrm{xmm}}$, temperature effects only & $0.799\pm0.007$ & $0.064\pm0.015$ & $0.17\pm0.01$ & 
   $-0.07_{-0.08}^{+0.10}$ & $-0.18_{-0.10}^{+0.11}$ & $-0.21_{-0.16}^{+0.20}$\\
$M_{500}^{\mathrm{xmm}}$, no time\-stamp correction & $0.826\pm0.004$ & $0.061\pm0.007$ & $0.15\pm0.01$ & 
   $-0.05_{-0.08}^{+0.10}$ & $-0.16_{-0.10}^{+0.11}$ & $-0.21_{-0.16}^{+0.20}$\\
  \hline\hline \label{tab:res}
  \end{tabular}
\end{table*}
Figure~\ref{fig:bias} shows the measured bias between the WL masses 
$M_{500}^{\mathrm{wl}}$ and $M_{500}^{\mathrm{xmm}}$ (including time\-stamp 
correction) for the I14 clusters. The bias is measured by averaging 
$\langle\log M^{\mathrm{xmm}}\!-\!\log M^{\mathrm{wl}}\rangle$ over
the suite of Monte Carlo simulations described in Sect.~\ref{sec:mc} 
that was used to obtain the $M_{500}^{\mathrm{xmm}}$ measurements.
The results are shown in Table~\ref{tab:res} and indicated by a dashed line 
and shading for the $1\sigma$ interval in Fig.~\ref{fig:bias}. Dashed lines 
and boxes at $M_{500}^{\mathrm{wl}}\!\leq\!10^{14.5}\,\mbox{M}_{\odot}$
and $M_{500}^{\mathrm{wl}}\!\geq\!10^{14.5}\,\mbox{M}_{\odot}$ show the bias 
for the thus defined low- and high-mass sub-samples.

For the eight clusters, we now find a pronounced bias of 
$b_{\mathrm{log}}\!=\!-0.08_{-0.08}^{+0.10}$, compared to 
$b_{\mathrm{log}}\!=\!0.02_{-0.8}^{+0.10}$ from \textit{Chandra}, using the
updated Monte Carlo method. For the low-mass sub-sample, 
$M_{500}^{\mathrm{xmm}}$ and $M_{500}^{\mathrm{WL}}$ are consistent
($b_{\mathrm{log}}\!=\!0.02_{-0.12}^{+0.16}$); while for the high-mass 
sub-sample, we measure $b_{\mathrm{log}}\!=\!-0.19_{-0.10}^{+0.11}$, 
i.e.\ $M_{500}^{\mathrm{xmm}}$ that are smaller than WL masses by a similar
amount as the $M^{\mathrm{Pl}}$ of vdL14 (cf.\ Fig.~\ref{fig:bias}). 

We repeat our analysis for a few modifications highlighting the
relative importance of various contributing factors:
First, we find that \textit{Chandra} masses, converted to the newer CalDB v4.2
and the $0.7$--$7$ keV band are systematically lower than for the V09a 
calibration and energy range. The \textit{Chandra}-only time\-stamp calibration
already accounts for $\sim\!30$~\% of the difference with \textit{XMM-Newton}:
$b_{\mathrm{log}}\!=\!-0.01_{-0.09}^{+0.10}$, a difference of 
$\Delta b_{\mathrm{log}}\!=\!-0.03$ (Table~\ref{tab:res}).
This result is consistent with the higher masses the V09a pipeline returns in 
the \citet{2014MNRAS.438...49R,2014MNRAS.438...62R} cross-calibration studies.
Conversely, omitting the time\-stamps correction moves up the 
$M_{500}^{\mathrm{xmm}}$, such that 
$b_{\mathrm{log}}\!=\!-0.05_{-0.08}^{+0.10}$ is less negative by 
$\Delta b_{\mathrm{log}}\!=\!0.03$. These comparisons demonstrate the 
importance of including the time\-stamp correction.

The $2$~\% difference the masses experience due to the different flux 
calibration of \textit{Chandra} and \textit{XMM-Newton} relates to a small,
but measurable effect in the logarithmic bias: Ignoring it, we find a slightly
milder bias of $b_{\mathrm{log}}\!=\!-0.07_{-0.08}^{+0.10}$ compared to the 
full conversion ($b_{\mathrm{log}}\!=\!-0.08_{-0.08}^{+0.10}$).

Considering the full mass range, the \textit{XMM-Newton} 
hydrostatic masses are $\sim\!20$~\% lower than the WL masses, while
\textit{Chandra} masses are consistent with the WL masses. This indicates that
if the $b_{\mathrm{lin}}\!=\!0.2$ linear hydrostatic bias in cluster 
simulations is correct, the effective area calibration of \textit{XMM-Newton} 
is consistent with being correct. But if looking at the high mass end, 
the conclusion is the opposite: \textit{Chandra} is consistent with the correct
calibration and $20$~\% hydro bias. The measurement uncertainties and the
unknown amount of Eddington bias in our small sample, however, preclude more 
quantitative conclusions.

\subsection{Mass-dependent bias with \textit{XMM-Newton}} \label{sec:massdepbias} 

Finally, we measure the mass-dependence of the bias as the difference 
$\Delta b_{\mathrm{log}}^{\mathrm{H-L}}$ between the logarithmic biases 
$b_{\mathrm{log}}$ for the high- and low-mass clusters. This observable is 
stable against changes to the details of the probability distribution 
modelling in the Monte Carlo algorithm. (Note that fitting 
$\log{(M^{\mathrm{xmm}}/M^{\mathrm{wl}})}$ as a function of
$M^{\mathrm{wl}}$ is not stable.) 

In I14, the hydrostatic mass exhibited the least significant mass-dependent 
bias of four tested mass observables. 
For the four more massive clusters, $b_{\mathrm{log}}$ is $\sim\!\!1\sigma$ 
different to the four less massive ones, as opposed to $\sim\!2\sigma$). 
We reproduce this result and measure 
$\Delta b_{\mathrm{log}}^{\mathrm{H-L}}\!=\!-0.20_{-0.16}^{+0.20}$ 
for \textit{Chandra} and
$\Delta b_{\mathrm{log}}^{\mathrm{H-L}}\!=\!-0.21_{-0.16}^{+0.20}$ 
for \textit{XMM-Newton} (Table~\ref{tab:res}). 

We interpret the observed mass-dependence of $b_{\mathrm{log}}$ as the 
superposition of 1.) physical effects, e.g. the stronger hydrostatic bias 
for high-mass clusters \citet{2014MNRAS.442..521S} predict analytically, and
2.) Eddington bias:
As \citet{2014arXiv1407.7868S} demonstrate, intrinsic scatter
in the abscissa mass leads to a mass-dependent bias when compared to an 
independent mass observable. 
Eddington bias is most severe in our case of a small sample size and a narrow
range in the underlying true mass. In principle, the statistically complete 
nature of the \emph{400d} cosmology (V09a) sample would allow for a rigorous
correction of such selection effects, once the WL follow-up has been completed.
For our given subsample, the Eddington bias and true mass-dependent mass bias 
cannot be disentangled. While we can provide much needed \textit{relative} 
cross-calibrations between X-ray and WL instruments/pipelines, selection 
effects preclude us from determining absolute calibrations for \textit{Chandra}
and \textit{XMM-Newton}. Moreover, selection biases also limit the direct
applicability of $\Delta b_{\mathrm{log}}^{\mathrm{H-L}}$ to other cluster samples.

\section{Translation to \textit{Planck} clusters} \label{sec:conclu}

\subsection{What did the \textit{Planck} collaboration measure?} \label{sec:pl}
\begin{figure}
 \includegraphics[width=8cm]{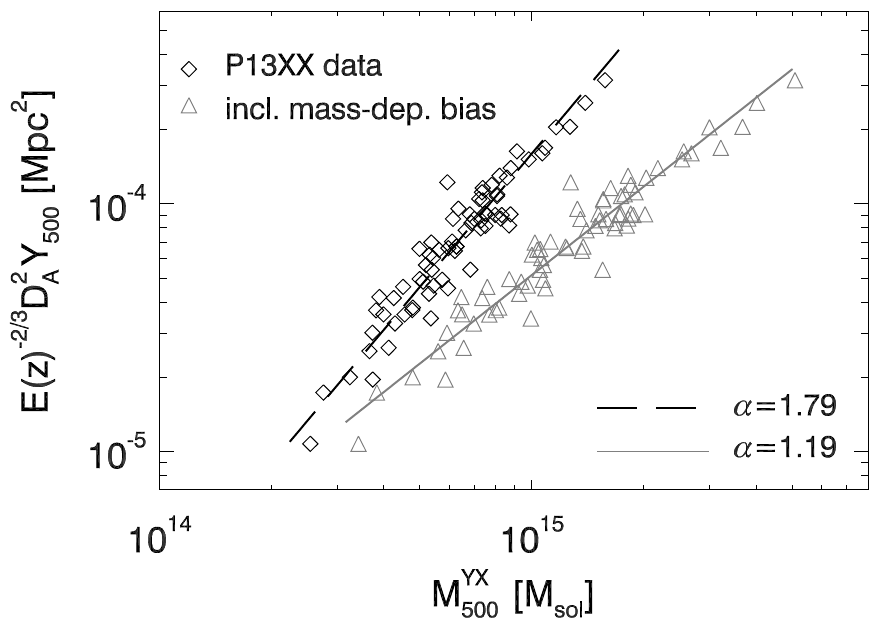}
 \caption{The P13XX calibration sample. Diamonds and the 
  long-dashed fit line show the SZ signal as a function of original P13XX 
  $Y_{\mathrm{X}}$ mass (compare their Fig.~A.1). Triangles and the solid fit
  line show rescaled masses, assuming an extreme case of a mass-dependent 
  hydrostatic bias.}
  \label{fig:planck}
\end{figure}

P13XX model the redshift-dependent abundance of clusters detected from the
\textit{Planck} catalogue of Sunyaev-Zeldovich sources 
\citep[P13XXIX]{2013arXiv1303.5089P}, covering the whole extragalactic sky.
The thermal SZ effect describes the inverse Compton scattering of CMB photons 
with ICM electrons, resulting in a distortion $Y_{\mathrm{SZ}}$ of the CMB
signal in the solid angle subtended by a galaxy cluster, proportional
to the integrated electron pressure.
All $189$ $S/N\!>\!7$ sources selected from the P13XXIX catalogue are confirmed
clusters of known redshift; the vast majority with spectroscopic redshifts.
The P13XXIX mass estimates $M^{\mathrm{Pl}}$ 
($M^{\mathrm{Y_{z}}}$ in P13XXIX) that enter the P13XX
calculation are the only, and crucial, piece of \textit{Planck} data 
P13XX use.  

Due to the large beam compared to the typical \textit{Planck} cluster size, 
the aperture size $\theta$, in which $Y_{\mathrm{SZ}}$ is integrated, 
is hard to determine from the SZ data itself.
P13XXIX rely on the additional $Y_{\mathrm{SZ}}(\theta)$ constraint
provided by the scaling of $Y_{\mathrm{SZ}}$ with an X-ray
mass proxy, $M^{Y_{\mathrm{X}}}_{500}$, to 
fix $\theta$ and 
calibrate the $M^{\mathrm{Pl}}$. By convention, $r_{\Delta}$ denotes a radius 
such that the mass $M_{\Delta}$ within it exceeds the critical density 
$\rho_{\mathrm{c}}(z)$ at redshift $z$ by a factor of $\Delta$.
The $M^{Y_{\mathrm{X}}}_{500}$ mass proxy is based on
$Y_{\mathrm{X}}\!=\!T_{\mathrm{X}}M_{\mathrm{gas}}$, which is the product of 
the ICM temperature $T_{\mathrm{X}}$ and the cluster gas mass 
$M_{\mathrm{gas}}$, measured from X-rays within $r_{500}$, and thus provides 
an X-ray analogue of $Y_{\mathrm{SZ}}$.

P13XX calibrate $M^{\mathrm{Pl}}$ on a validation sub-sample of $71$ clusters
observed with \textit{XMM-Newton}, i.e.\ they derive the best-fit 
$Y_{\mathrm{SZ,500}}$--$M^{Y_{\mathrm{X}}}_{500}$ relation.
In turn, $M^{Y_{\mathrm{X}}}_{500}$ was calibrated on a sample of local, 
relaxed clusters whose ``true'' masses could be measured using X-ray 
observations and assuming hydrostatic equilibrium \citep{2010A&A...517A..92A}.
All EPIC instruments were used, with the 
pn/MOS normalisation as a free parameter. Spectra were fitted in the
$0.3$--$10$ keV energy band (M.\ Arnaud; priv.\ comm.).
It is via this ladder of mass proxies that the hydrostatic mass bias is
inherited onto $M^{\mathrm{Pl}}$, appearing in the 
$Y_{\mathrm{SZ,500}}$--$M^{Y_{\mathrm{X}}}_{500}$ relation that summarises
the calibration process (Eq.~A.8 of P13XX). P13XX considered a flat prior of 
$0.7\!\!<\!\!(1\!\!-\!b_{\mathrm{lin}})\!\!<\!\!1$, 
but any additional systematic effect in the calibration chain would mimic
a spurious ``hydrostatic'' bias.

\subsection{Comparison to \textit{Planck} and vdL14 samples} \label{sec:comparison}

The mean WL mass of the I14 
high-mass sub-sample is $4.9\cdot10^{14}\,\mbox{M}_{\odot}$.
The typical P13XX cluster mass, defined by their mass pivot
$\sim\!6\cdot10^{14}\,\mbox{M}_{\odot}$, falls into the mass range probed by
the I14 high-$M^{\mathrm{wl}}$ range, even although the mass bias is not 
included. Therefore, for the relevant P13XX mass range,our result of 
$b_{\mathrm{log,H}}\!=\!0.20_{-0.16}^{+0.17}$ 
agrees with the $1\!-\!b_{\mathrm{lin}}\!\approx\!0.4$ that would
reconcile cosmological constraints derived from \textit{Planck} cluster counts
(P13XX) and primary CMB anisotropies (P13XVI).

The high-mass end of the I14 sample also overlaps with the vdL14 sample.
Using the $M_{500}^{\mathrm{xmm}}$ for the I14 clusters instead of 
\textit{Chandra} masses, we also find better agreement to the vdL14 measurement 
of $\langle M^{\mathrm{Pl}}\!/M^{\mathrm{wl}}\rangle\!=\!0.688\pm0.072$
for a subset of P13XX clusters.
However, such comparisons are limited by the small number statistics of our
sample, hence caution is necessary when interpreting these 
results.\footnote{The difference in cosmologies between P13XX and vdL14 on the
one hand (flat universe with matter density $\Omega_{\mathrm{m}}\!=\!0.3$ and
Hubble parameter $H_{0}\!=\!70\,\mbox{km}\,\mbox{s}^{-1}\,\mbox{Mpc}^{-1}$) 
and I14 and this work the other hand (the same, but
$H_{0}\!=\!72\,\mbox{km}\,\mbox{s}^{-1}\,\mbox{Mpc}^{-1}$) adds a factor of
$70/72$ to convert $\textit{Planck}$ masses to our cosmology.}

Complications arise from the different energy range used for \textit{Planck} 
and the temporal variability of X-ray calibrations.
Our results for the cases with and without time\-stamp correction
(Table~\ref{tab:res}) tell us, however, that the impact of those systematics
is rather small, with $\Delta b_{\mathrm{lin}}\!\la\!0.05$.

\subsection{How much can X-ray calibration bias have influenced the P13XX results?}

\subsubsection{From \textit{Planck} pre-calibration to calibration}

We attempt to estimate how an additional bias 
$b^{\mathrm{xcal}}_{\mathrm{lin}}$ arising from the $\textit{XMM-Newton}$ 
calibration relative to \textit{Chandra} will influence the overall bias 
measured by P13XX. We emphasise that we do not know or assume which, if any, 
satellite calibration is correct.
The ``pre-calibration" from $20$ relaxed clusters \citep{2010A&A...517A..92A} 
determines the normalisation $10^{B}$ and slope $\beta$ of a scaling relation
\begin{equation}
E^{-2/3}(z)\left[\frac{Y_{\mathrm{X}}}{2\cdot10^{14}\,\mbox{M}_{\odot}\,\mbox{keV}}\right]
=10^{B}\cdot\left[\frac{M_{500}^{\mathrm{HE}}}{6\cdot10^{14}\,\mbox{M}_{\odot}}\right]^{\beta}
\label{eq:a10}
\end{equation}
between the $Y_{\mathrm{X}}$ and hydrostatic masses $M_{500}^{\mathrm{HE}}$
measured with \textit{XMM-Newton}. The evolution factor 
$E(z)\!=\!H(z)/H(z\!=\!0)$ depends on cosmology via the Hubble parameter $H(z)$.

In Eq.~(\ref{eq:a10}), $M_{500}^{\mathrm{HE}}$ scales roughly as 
$T_{\mathrm{X}}^{3/2}$ \citep[e.g.,][]{2012MNRAS.422.1999K}, through the 
measurement at $r_{500}$.
If $q\!=\!T_{\mathrm{XMM}}/T_{\mathrm{CXO}}$ for the typical 
\citet{2010A&A...517A..92A} cluster, hydrostatic masses are biased
$M_{500}^{\mathrm{HE}}\!\rightarrow\!q^{\delta}M_{500}^{\mathrm{HE}}$, with
$\delta\!\approx\!1.5$. 
Similarly, $Y_{\mathrm{X}}$ depends on $T_{\mathrm{X}}$ via the measurement
of the gas mass $M_{\mathrm{gas}}$ within $r_{500}$: 
We have $r_{500}\!\propto\!M_{500}^{1/3}$. 
If $M_{500}\!\propto\!T_{\mathrm{X}}^{3/2}$ upon a change in $T_{\mathrm{X}}$,
then $r_{500}\!\propto\!(T_{\mathrm{X}}^{3/2})^{1/3}\!=\!T_{\mathrm{X}}^{1/2}$.
Because $M_{\mathrm{gas}}(<\!r)$ increases linearly with $r$ in a given 
cluster\footnote{If the cluster is isothermal, and 
$\rho_{\mathrm{gas}}\!\propto\!r^{-2}$, as
motivated by assuming the standard $\beta\!=\!2/3$ in the $\beta$ model for the
gas density \citep{1978A&A....70..677C}, then the 3D mass within a radius $R$ is
$M(<\!R)\!=\!\int_{0}^{R}{\rho_{\mathrm{gas}}(r)\,\mathrm{d}V} \propto 
\int_{0}^{R}{r^{-2}\,r^{2}\,\mathrm{d}r}\!=\!R$.} 
it follows $M_{\mathrm{gas,500}}\!\propto\!T_{\mathrm{X}}^{1/2}$ upon a change in 
$T_{\mathrm{X}}$. Indeed, we measure $M_{\mathrm{gas,500}}$ to be affected
as $q^{0.5-0.6}$ to by a relative temperature change $q$, using the
V09a gas density model for the I14 clusters.
Hence, we have $Y_{\mathrm{X}}\!\rightarrow\!q^{\gamma}Y_{\mathrm{X}}$ with an
exponent $\gamma\!\approx\!1.5$. Thus $T_{\mathrm{X}}$ re-calibration 
affects Eq.~(\ref{eq:a10}) like:
\begin{equation}
q^{\gamma}Y_{\mathrm{X}}\!\propto\!\left[q^{\delta}M_{500}^{\mathrm{HE}}\right]^{\beta}\Leftrightarrow
Y_{\mathrm{X}}\!\propto\!q^{\beta\delta-\gamma}\left[M_{500}^{\mathrm{HE}}\right]^{\beta}.
\end{equation}
For a (residual, unaccounted) temperature bias $q$, the mass
proxy $M_{500}^{Y_{\mathrm{X}}}$ will be biased by a factor 
$C\!=\!q^{\beta\delta-\gamma}$. This factor propagates into
the main P13XX scaling relation, connecting the masses 
$M_{500}^{Y_{\mathrm{X}}}$ to $Y_{500}$ instead of $Y_{\mathrm{X}}$:
\begin{equation}
E^{-2/3}(z)\left[\frac{D_{\mathrm{A}}^{2}Y_{\mathrm{SZ,500}}}{10^{-4}\,\mbox{Mpc}^{2}}\right]=10^{A}
q^{\alpha\delta-\gamma}\cdot\left[\frac{M_{500}^{Y_{\mathrm{X}}}}{6\cdot10^{14}\,\mbox{M}_{\odot}}\right]^{\alpha}.
\label{eq:mcalib}
\end{equation}
Here, $D_{\mathrm{A}}$ denotes the angular diameter distance. Because
$Y_{\mathrm{X}}$ is theoretically expected to be proportional to 
$Y_{\mathrm{SZ}}$, we identified $\alpha\!=\!\beta$ in Eq.~(\ref{eq:mcalib}).

\subsubsection{Results for temperature re-calibration}

The \citet{2010A&A...517A..92A} clusters used in the \textit{Planck}
pre-calibration show an average \footnote{In principle, the
temperature recalibration should be applied to individual clusters.
This would alter the slope $\beta$ in Eq.~(\ref{eq:a10})
in a similar way as the mass-dependent mass bias discussed below.} 
$k_{\mathrm{B}}T_{\mathrm{XMM}}\!\approx\!5\pm2\,\mbox{keV}$
\citep{2007A&A...474L..37A,2010A&A...511A..85P}.
Following Eq.~(\ref{eq:s14}), the S14 conversion for the combined 
\textit{XMM-Newton} instruments, \textit{Chandra} temperatures for these 
clusters would be lower by a factor of $q\!=\!0.84_{-0.03}^{+0.05}$.

Using $\alpha\!=\!1.79\pm0.06$ from P13XX, and
$\gamma\!=\!1.5\pm0.3$ and $\delta\!=\!1.5\pm0.3$ 
(i.e.\ allowing for broad uncertainties in both), 
we find the normalisation of Eq.~(\ref{eq:mcalib}) to be reduced by 
a factor of $C\!=\!0.81\pm0.09$.

\subsubsection{Breaking the size-flux degeneracy} \label{sec:sizeflux}

The exact algorithm by which P13XXIX combine \textit{Planck} measurements with
Eq.~(\ref{eq:mcalib}) has yet to be published. However, using 
$\theta_{500}\!=\!\left(3M_{500}/[4\pi\rho_{\mathrm{c}}D_{\mathrm{A}}^{3}]\right)^{1/3}$,
one can easily convert Eq.~(\ref{eq:mcalib}) into a scaling relation in terms 
of an aperture scale $\theta_{500}$, i.e.: 
$Y_{\mathrm{SZ}}\!\propto\!\theta_{500}^{3\alpha}$. 
The intersection of this relation with the size--flux degeneracy modelled as 
$Y_{\mathrm{SZ}}^{\mathrm{obs}}\!\propto\!\theta^{\lambda}$ yields a point
($\theta_{\times}$,$Y_{\times}$), that can in turn be used to 
compute an SZ mass $M_{\mathrm{Pl}}\!\propto\!\theta_{\times}^{3}$. 
Thus, the degeneracy is broken. 
How is this $M_{\mathrm{Pl}}$ affected if the normalisation of 
Eq.~(\ref{eq:mcalib}) changes by a factor $C$? 
We geometrically infer the changes in the intersection point and final mass as:
\begin{eqnarray}
\log{(Y'_{\times}/Y_{\times})}&=&\left[-\lambda/(\lambda\!-\!3\alpha)\right]\cdot\log{C}\\
\log{C_{\mathrm{fin}}}\!=\!\log{(M'_{\mathrm{Pl}}/M_{\mathrm{Pl}})}&=&
\left[-3\alpha\lambda/(\lambda\!-\!3\alpha)\right]\cdot\log{C}.
\end{eqnarray} \label{eq:sizeflux}
From Fig.~4 of P13XXIX, we read that the 
$Y_{\mathrm{SZ}}^{\mathrm{obs}}$--$\theta$ relation is linear, 
so $\lambda\!=\!1$. With $C\!=\!0.81_{-0.03}^{+0.05}$ 
from above, we find that cluster masses would be biased low by a factor 
$C_{\mathrm{fin}}\!=\!0.78_{-0.07}^{+0.10}$ due to the temperature 
calibration. Thus, if the \textit{Chandra} calibration was correct, the need 
for a hydrostatic mass bias of more than the $\sim\!20$~\% favoured by 
simulations would be eased. Alternatively, if the \textit{XMM-Newton} 
calibration was correct, evidence for stronger departures from 
hydrostatic equilibrium would persist.

We note that the ``hydrostatic'' bias $b_{\mathrm{lin}}$ that P13XX consider 
is meant to include instrument calibration effects:
$1\!-\!b_{\mathrm{lin}}\!=\!(1\!-\!b_{\mathrm{lin}}^{\mathrm{hyd}})(1\!-\!b_{\mathrm{lin}}^{\mathrm{xcal}})
\!\approx\!(1\!-\!b_{\mathrm{lin}}^{\mathrm{hyd}}\!-\!b_{\mathrm{lin}}^{\mathrm{xcal}})$.
Nevertheless even a partially unaccounted calibration bias would contribute
some of the apparent mass discrepancy.
The point of this exercise lies not in suggesting that the \textit{Planck}
discrepancy is caused by the X-ray calibration. Rather it should serve to
demonstrate how such effects can not only fold through but even become
amplified in a multi-step calibration.

\subsubsection{Inclusion of the mass-dependent bias}

The above calculations treat the case of a potential residual temperature 
calibration offset in the \textit{Planck} calibration. To this end, we assume
the hydrostatic mass bias to be taken into account and well represented by the 
P13XX baseline value of $(1\!-\!b_{\mathrm{hyd}})\!=\!0.8_{-0.1}^{+0.2}$.

But it is instructive to include a mass-dependent hydrostatic mass bias,
as suggested by I14 and Fig.~\ref{fig:bias}. Because we are interested in 
extreme cases, we assume that the best-fit 
$b_{\mathrm{log}}(M_{500}^{\mathrm{xmm}})\!=\!-0.346\cdot(E(z) M_{500}^{\mathrm{xmm}}/2.44\cdot\!10^{14}\,M_{\odot}) -0.111$
is purely physical (departure from hydrostatic equilibrium).
We emphasise this is not the case: As detailed in Sect.~\ref{sec:massdepbias}, 
not all of the mass-dependence is physical, but an unknown
fraction is caused by selection effects (Eddington bias). 

Figure~\ref{fig:planck} shows how a mass-dependent mass bias differentially
stretches the mass range occupied by the \textit{Planck} calibration clusters.
In our extreme scenario, masses for all clusters are higher after accounting
for $b_{\mathrm{log}}$ (triangles) than before (diamonds), but most so for the
most massive ones. Consequentially the slope of Eq.~(\ref{eq:mcalib}) needs to
be corrected from P13XX's $\alpha\!=\!1.79\pm0.06$ to a lower value of 
$\alpha\!=\!1.19\pm0.04$.\footnote{Observations of the $Y$--$M$ relation have
yet to reach an accuracy that would such constrain the mass-dependency of the
hydrostatic bias. While \citet{2014arXiv1404.7103B} and 
\citet{2014arXiv1406.2800C} report low best-fit $Y$--$M$ slopes consistent 
with $\alpha\!\approx\!1.2$, \citet{2014arXiv1407.7520L} find a
slope steeper than the self-similar value of $5/3$.}

Interestingly, a flatter $Y$--$M$ slope largely cancels out the temperature
re-calibration effect seen in Sect.~\ref{sec:sizeflux}. With 
$\alpha\!=\!1.19\pm0.04$, we arrive at a factor of $C\!=\!0.95_{-0.05}^{+0.08}$ 
in Eqs.~(\ref{eq:mcalib}) and final \textit{Planck} masses different by a 
factor of $C_{\mathrm{fin}}\!=\!0.94_{-0.07}^{+0.11}$. We conclude that 
inclusion of a mass-dependent hydrostatic bias that grows more negative with
mass cannot increase the final calibration offset. The better, still partial, 
alleviation of the \textit{Planck} cluster counts--CMB discrepancy
is achieved from X-ray calibration effects alone.

\section{Summary and Outlook} \label{sec:conclusion}

Starting from the recent \citet{2014arXiv1404.7130S} comparative study of ICM
temperatures measured with \textit{Chandra} and \textit{XMM-Newton}, we revisit
the bias between WL and hydrostatic masses from \citet{2014A&A...564A.129I}.
We find:

\begin{description}
\item[1.] Because of different uncertainties in the effective area calibration, 
hydrostatic masses for the I14 clusters would have been measured to be 
$\sim\!15$--$20$~\% lower, had the clusters been observed with 
\textit{XMM-Newton} instead of \textit{Chandra}. The measured calibration bias
depends on the sample, but can be transferred to clusters of similar mass 
($10^{14}$--$10^{15}\,\mbox{M}_{\odot}$).
\item[2.] \textit{XMM-Newton} masses for the most massive I14 clusters are 
lower than WL masses by $\sim\!35$~\%.
\item[3.] 
Assuming a true hydrostatic bias of $b_{\mathrm{lin}}^{\mathrm{hyd}}\!=\!0.2$,
our results for the whole mass range indicate that the calibration of the 
energy dependence of the effective area of the \textit{XMM-Newton} EPIC 
instruments in the $0.6$--$10.0\,\mbox{keV}$ band is rather accurate. 
In the high mass range the data however indicate that \textit{Chandra} 
calibration is more accurate. Given the uncertainties these results are
not significant.
\end{description}

In addition, we consider the \textit{Planck} clusters and find: 

\begin{description}
\item[4.] Hence, consistent with vdL14, a bias of
$(1\!-\!b_{\mathrm{lin}}^{\mathrm{hyd}}\!-\!b_{\mathrm{lin}}^{\mathrm{xcal}})\!\approx\!0.4$
for the rather massive P13XX clusters seems plausible.
\item[5.] If there was a residual calibration bias $q$ in the 
$T_{\mathrm{XMM}}$ measurements on which the \textit{Planck} analysis is based,
the normalisation of the P13XX $Y_{\mathrm{SZ}}$--$M^{\mathrm{Y_{X}}}$
calibration would be affected as $C\!=\!q^{\sim\!1.2}$. 
We show that the mass bias further amplifies when propagated into the SZ 
masses. Without accounting for calibration uncertainties, a mass bias of up to 
$20$~\% is plausible. We do not claim that this is the case for 
\textit{Planck}. However, a small, residual bias would amplify in the same way.
Pointing to the S14 result that calibration alone cannot explain the discrepant
cosmological parameters of P13XVI and P13XXIX, we conclude that a possible 
contribution would ease the discrepancy and allow for a true hydrostatic bias
consistent with simulations.
\item[6.] A hydrostatic bias increasing with mass counteracts the amplification
of a calibration bias. 
\end{description}

Our results are consistent with the WL/X-ray mass biases recently reported by
\citet{2014arXiv1405.7876D}, comparing CLASH WL mass profiles to those obtained
with \textit{Chandra} and \textit{XMM-Newton}.
\citet{2014arXiv1405.7876D} found their $T_{\mathrm{XMM}}/T_{\mathrm{CXO}}$ and
$M^{\mathrm{XMM}}\!/M^{\mathrm{WL}}$ to depend on the integration radius;
suggesting soft X-ray scattering as a cause for the calibration offset.
\citet{2014arXiv1405.7876D} study mostly cool core clusters.
Since S14 find that the $T_{\mathrm{X}}$ bias depends on $T_{\mathrm{X}}$, 
this could explain why they find less bias in the cooler centres. 
The radial dependence could at least partly be due to a secondary correlation: 
at the radius where the cluster temperature is typically hottest, the largest 
discrepancy between \textit{Chandra} and \textit{XMM-Newton} is found.

Cluster mass calibrations still bear considerable uncertainties not only
between the main techniques (X-ray, lensing, SZ, galaxy-based), but also
within techniques, i.e\ for different instruments and calibration and methods.
Thorough cross-calibration of different instruments and techniques, as already
performed by 
\citet{2010A&A...523A..22N,2014arXiv1404.7130S,2014MNRAS.438...49R,2014MNRAS.438...62R}
for X-rays are the necessary way forward. Recent comparisons of WL masses to
both \textit{XMM-Newton} and \textit{Chandra} include 
\citet{2013ApJ...767..116M, 2014arXiv1405.7876D}, and 
\citet{2014arXiv1406.6831M}. We notice that \citet{2014arXiv1406.6831M} find 
temperature discrepancies between \textit{XMM-Newton} and \textit{Chandra} 
similar to S14, but consistent hydrostatic masses from both satellites. 
More overlap between clusters with X-ray and WL data would be necessary to 
define mass standards against which other surveys could then be gauged.

Recently, \citet{2014arXiv1407.7868S,2014arXiv1407.7869S} compared several of
the larger current WL and \textit{XMM-Newton} and \textit{Chandra} X-ray 
samples, emphasising how intrinsic and measurement scatter can induce scaling
relation biases. \citet{2014arXiv1407.7868S} confirm that compared to 
simulated clusters WL masses are biased low by $\sim\!10$~\% and hydrostatic
masses by $\sim\!20$--$30$~\%. However, these authors find literature masses
\emph{from the same observable}, X-ray or WL, can differ up to $40$~\%
for the same cluster, impeding an absolute calibration. 
\citet{2014arXiv1407.7869S} extend the analysis to the \textit{Planck} 
clusters, whose absolute mass calibration is likewise affected. They find
scatter in the calibration scaling relation
to invoke a mass-dependent bias in the \textit{Planck} masses.

Here, the \emph{400d} cluster sample provides \emph{relative} 
calibrations between the different instruments and methods. Once the WL 
follow-up has been completed, we will be able to disentangle the physical 
mass-dependent mass bias from selection effects, and
provide \emph{absolute} calibrations.

The advent of larger SZ samples for scaling relation studies 
\citep[e.g.,][]{2014arXiv1404.7103B,2014arXiv1406.2800C,2014arXiv1407.7520L},
and foremost the all-sky P13XXIX offers the possibility to include a 
complementary probe and clusters at higher redshift. 
For future high precision cluster experiments, e.g., \textit{eROSITA}
\citep{2010SPIE.7732E..23P,2012arXiv1209.3114M,2012MNRAS.422...44P}
or \textit{Euclid} \citep{2011arXiv1110.3193L,2012arXiv1206.1225A} 
the absolute X-ray observable--mass calibration needs to be improved further.

\bibliographystyle{mn2e}
\bibliography{recal_resubmit}

\section*{acknowledgements}
We thank the the \textit{Planck} collaboration for making
available the calibration sample in the SZ cluster database
(\texttt{http://szcluster-db.ias.u-psud.fr}). 
We thank the referee for helpful comments. \color{black}
HI would like to thank D.\ Applegate for a helpful discussion.
HI acknowledges support from European Research Council grant MIRG-CT-208994 
and Philip Leverhulme Prize PLP-2011-003.
JN acknowledges a PUT 246 grant from Estonian Research Council.
RJM is supported by a Royal Society University Research Fellowship.
THR acknowledges support from the German Research Association (DFG)
through Heisenberg grant RE 1462/5 and through the Transregional
Collaborative Research Centre TRR33 ``The Dark Universe'' (project B18).
GS and THR acknowledge DFG grant RE 1462/6.

\label{lastpage}

\end{document}